\documentclass[apjl,onecolumn]{emulateapj}

\usepackage{natbib}

\newcommand{\dfrac}[2]{{\displaystyle \frac{#1}{#2}}  }

\newcommand{\eqref}[1]{(\ref{#1})}

\def\lesssim{\mathrel{\hbox{\rlap{\hbox{\lower4pt\hbox{$\sim$}}}\hbox{$<$}}}}
\def\gtrsim{\mathrel{\hbox{\rlap{\hbox{\lower4pt\hbox{$\sim$}}}\hbox{$>$}}}}

\slugcomment{ApJL, in press}

\shortauthors{Muto et al.}
\shorttitle{Spiral in the Disk of SAO 206462}

\begin{document}

\title{Discovery of Small-Scale Spiral Structures in the Disk of 
SAO 206462 (HD 135344B)\footnotemark[*]
    : Implications for the Physical State of
    the Disk from Spiral Density Wave Theory}

\author{T. Muto\altaffilmark{1,35,36}, 
C. A. Grady\altaffilmark{2,3,4}, 
J. Hashimoto\altaffilmark{5}, 
M. Fukagawa\altaffilmark{6},
J. B. Hornbeck\altaffilmark{7}, 
M. Sitko \altaffilmark{8,9,10},
R. Russell \altaffilmark{10,11}, 
C. Werren \altaffilmark{9,10},  
M. Cur\'e\altaffilmark{12}, 
T. Currie\altaffilmark{3}, 
N. Ohashi\altaffilmark{13,14}, 
Y. Okamoto\altaffilmark{15}, 
M. Momose\altaffilmark{15}, 
M. Honda\altaffilmark{16}, 
S. Inutsuka\altaffilmark{17},  
T. Takeuchi\altaffilmark{1}, 
R. Dong\altaffilmark{18}, 
L. Abe\altaffilmark{19},
W. Brandner\altaffilmark{20},
T. Brandt\altaffilmark{18},
J. Carson\altaffilmark{21},
S. Egner\altaffilmark{14},
M. Feldt\altaffilmark{20},
T. Fukue\altaffilmark{5},
M. Goto\altaffilmark{20},
O. Guyon\altaffilmark{14},
Y. Hayano\altaffilmark{14},
M. Hayashi\altaffilmark{22,37},
S. Hayashi\altaffilmark{14},
T. Henning\altaffilmark{20},
K. W. Hodapp\altaffilmark{23},
M. Ishii\altaffilmark{14},
M. Iye\altaffilmark{5},
M. Janson\altaffilmark{18},
R. Kandori\altaffilmark{5},
G. R. Knapp\altaffilmark{18},
T. Kudo\altaffilmark{14},
N. Kusakabe\altaffilmark{5},
M. Kuzuhara\altaffilmark{5,24},
T. Matsuo\altaffilmark{25},
S. Mayama\altaffilmark{26},
M. W. McElwain\altaffilmark{3},
S. Miyama\altaffilmark{5},
J.-I. Morino\altaffilmark{5},
A. Moro-Martin\altaffilmark{18,27},
T. Nishimura\altaffilmark{14},
T.-S. Pyo\altaffilmark{14},
E. Serabyn\altaffilmark{28},
H. Suto\altaffilmark{5},
R. Suzuki\altaffilmark{29},
M. Takami\altaffilmark{13},
N. Takato\altaffilmark{14},
H. Terada\altaffilmark{14},
C. Thalmann\altaffilmark{30},
D. Tomono\altaffilmark{14},
E. L. Turner\altaffilmark{18,31},
M. Watanabe\altaffilmark{32},
J. P. Wisniewski\altaffilmark{33},
T. Yamada\altaffilmark{34},
H. Takami\altaffilmark{14},
T. Usuda\altaffilmark{14},
and
M. Tamura\altaffilmark{5}}

\footnotetext[*]{Based on data collected at the Subaru Telescope,
which is operated by the National Astronomical Observatory of Japan.}

\altaffiltext{1}{Tokyo Institute of Technology, 2-12-1 Ookayama,
Meguro, Tokyo 152-8551, Japan}
\altaffiltext {2}{Eureka Scientific, 2452 Delmer, Suite 100, Oakland CA
96002, USA}  
\altaffiltext {3}{ExoPlanets and Stellar Astrophysics Laboratory, Code
667, Goddard Space Flight Center, Greenbelt, MD 20771 USA} 
\altaffiltext {4}{Goddard Center for Astrobiology, Code 667, Goddard
Space Flight Center, Greenbelt, MD 20771 USA}
\altaffiltext {5}{National Astronomical Observatory of Japan, 2-21-1
Osawa, Mitaka, Tokyo 181-8588, Japan} 
\altaffiltext {6}{Department of Earth and Space Science, Graduate
School of Science, Osaka University, 1-1, Machikaneyama, Toyonaka, Osaka
560-0043, Japan} 
\altaffiltext {7} {Department of Physics and Astronomy, University of
Louisville, Louisville, KY 40292, USA} 
\altaffiltext {8}{Space Science Institute, 4750 Walnut St., Suite 205,
Boulder, CO 80301, USA} 
\altaffiltext{9}{Department of Physics, University of Cincinnati,
Cincinnati, OH 45221-0011, USA} 
\altaffiltext{10}{ Visiting Astronomer, NASA Infrared Telescope
Facility, operated by the University of Hawaii under contract to NASA}
\altaffiltext{11} {The Aerospace Corporation, Los Angeles, CA 90009, USA}
\altaffiltext {12} {Departamento de F\'isica y Astronom\'ia, Universidad de
Valpara\'iso, Avda. Gran Breta\~na 1111, Casilla 5030, Valpara\'iso, Chile } 
\altaffiltext {13} {Institute of Astronomy and Astrophysics, Academia
Sinica, P.O. Box 23-141, Taipei 106, Taiwan} 
\altaffiltext{14} {Subaru Telescope, 650 North A'ohoku Place, Hilo, HI
96720, USA} 
\altaffiltext {15} {College of Science, Ibaraki University, 2-1-1
Bunkyo, Mito, Ibaraki 310-8512, Japan }
\altaffiltext {16} {Department of Information Science, Kanagawa
University, 2946 Tsuchiya, Hiratsuka, Kanagawa 259-1293, Japan} 
\altaffiltext {17} {Department of Physics, Nagoya University, Furo-cho,
Chikusa-ku, Nagoya, Aichi, 464-8602, Japan}
\altaffiltext {18} {Department of Astrophysical Sciences, 
Princeton University, NJ08544, USA}
\altaffiltext{19} {Laboratoire Lagrange, UMR7293, Universit\'e de
Nice-Sophia Antipolis, CNRS, Observatoire de la C\^ote d'Azur, 06300
Nice, France}
\altaffiltext{20} {Max Planck Institute for Astronomy, Heidelberg, Germany}
\altaffiltext{21} {Department of Physics and Astronomy, College of
Charleston, 58 Coming St., Charleston, SC 29424, USA}
\altaffiltext{22} {Department of Astronomy, The University of Tokyo,
Hongo 7-3-1, Bunkyo-ku, Tokyo 113-0033, Japan}
\altaffiltext{23} {Department of Earth and Planetary Science, The
University of Tokyo, Hongo 7-3-1, Bunkyo-ku, Tokyo 113-0033, Japan}
\altaffiltext{24} {Institute for Astronomy, University of Hawaii, 640
North A'ohoku Place, Hilo, HI 96720, USA}
\altaffiltext{25} {Department of Astronomy, Kyoto University,
Kitashirakawa-Oiwake-cho, Sakyo-ku, Kyoto, 606-8502, Japan}
\altaffiltext{26} {The Graduate University for Advanced
Studies(SOKENDAI), Shonan International Village, Hayama-cho, Miura-gun, 
Kanagawa 240-0193, Japan}
\altaffiltext{27} {Departamento de Astrof\'isica, CAB (INTA-CSIC), 
Instituto Nacional de T\'ecnica Aeroespacial, Torrej\'on de Ardoz,
28850, Madrid, Spain}
\altaffiltext{28} {Jet Propulsion Laboratory, California Institute of
Technology, Pasadena, CA 91109, USA}
\altaffiltext{29} {TMT Observatory Corporation, 1111 South Arroyo
Parkway, Pasadena, CA 91105, USA}
\altaffiltext{30} {Astronomical Institute "Anton Pannekoek", University
of Amsterdam, Science Park 904, 1098 XH Amsterdam, The Netherlands}
\altaffiltext{31} {Kavli Institute for the Physics and Mathematics of
the Universe, Todai Institutes for Advanced Study, the University of
Tokyo, Kashiwa, Japan 277-8583 (Kavli IPMU, WPI)}
\altaffiltext{32} {Department of Cosmosciences, Hokkaido University,
Sapporo 060-0810, Japan}
\altaffiltext{33} {Department of Astronomy, University of Washington,
Box 351580 Seattle, Washington 98195, USA}
\altaffiltext{34} {Astronomical Institute, Tohoku University, Aoba,
Sendai 980-8578, Japan}
\altaffiltext{35} {JSPS Research Fellow}
\altaffiltext{36} {As of April 2012: Division of Liberal Arts, Kogakuin
University, 1-24-2, Nishi-Shinjuku, Shinjuku-ku, Tokyo, 163-8677, Japan}
\altaffiltext{37} {as of April 2012: National Astronomical Observatory
of Japan}

\email{muto@geo.titech.ac.jp}

\begin{abstract}

 We present high-resolution, $H$-band, imaging observations, collected with
 Subaru/HiCIAO, of the scattered light from
 the transitional disk around SAO 206462 (HD 135344B). 
 Although previous sub-mm imagery suggested the existence of the 
 dust-depleted cavity at $r \leq 46\ \mathrm{AU}$, our observations reveal 
 the presence of scattered light components as close as 
 $0 \farcs 2\ (\sim 28 \ \mathrm{AU})$ from the star. 
 Moreover, we have discovered two small-scale spiral structures lying
 within $0 \farcs 5 \ (\sim 70\ \mathrm{AU})$. 
 We present models for the spiral structures  
 using the spiral density wave theory, and derive a disk aspect ratio
 of $h \sim 0.1$, which is consistent with previous sub-mm observations.  
 This model can potentially give estimates of the temperature
 and rotation profiles of the disk based on dynamical processes, 
 independently from sub-mm observations.   
 It also predicts the evolution of the spiral structures, which
 can be observable on timescales of 10-20 years, providing 
 conclusive tests of the model.  
 While we cannot uniquely identify the origin of these spirals, 
 planets embedded in the disk may be capable of exciting the observed 
 morphology. 
 Assuming that this is the case, we can make predictions on the
 locations and, possibly, the masses of the unseen planets. 
 Such planets may be detected by future multi-wavelengths observations.  

\end{abstract} 


 \section {Introduction} 
 \label{sec:intro}

Dynamical processes in protoplanetary disks such as turbulence or 
disk-planet interaction are 
important in understanding physical condition and  
evolution of disks, and planet formation processes.  
High resolution, direct imaging observations of
circumstellar/protoplanetary disks 
can reveal non-axisymmetric structures, providing 
insight into such dynamical processes \citep[e.g.,][]{hash11}. 

Recent observations have identified a class of protoplanetary disks 
harboring tens of AU-scale holes/gaps at their centers: 
the so-called transitional disks. 
One well-studied system of that class is 
the rapidly rotating Herbig F star, SAO 206462 
\citep[HD 135344B, F4Ve, $d=142\pm27 \ \mathrm{pc}$,
$M=1.7^{+0.2}_{-0.1} M_\odot$,][]{mull11}.
The observations of CO line profiles \citep{dent05,pon08,lyo11} 
and stellar rotation \citep{mull11} 
consistently indicate an almost face-on geometry 
($i \sim 11^{\circ}$).  
The gap in the disk was predicted from the 
infrared (IR) spectral energy distribution \citep[SED,][]{bro07}, 
and was subsequently imaged in sub-mm dust continuum at 
$\sim 0 \farcs 5 \times 0 \farcs 25$ resolution 
\citep[][]{bro09}.
\citet{and11} estimate the gap radius to be $\sim 46 \ \mathrm{AU}$ and
the surface density within the gap to be $10^{-5.2}$ times smaller than
that extrapolated from the outer disk.
The gas in Keplerian motion surrounding the gap region  
is also imaged by CO lines \citep{lyo11}.
The CO rovibrational line observations \citep{pon08} 
and [OI] spectral line observations \citep{vdP08} 
 indicate the presence of a gas disk in the vicinity 
 (several AU-scale) of the star. 
SED modeling \citep{gra09} and NIR interferometry \citep{fed08} 
indicate the presence of an inner dust belt, which is 
temporally variable \citep{sit11} and not coplanar with the outer disk  
 (Benisty 2011, private communication). 
New imaging with high spatial resolution and sensitivity is required  
to understand the inner structures of the disk. 
 The outer portions of gaps can now be resolved 
using 8-10 m ground-based telescopes at near infrared (NIR) wavelengths
\citep[e.g.,][for LkCa 15]{thal10}. 

In this Letter, we present $H$-band polarized intensity ($PI$) 
observations of the disk of SAO 206462 down to  
$r \sim 0 \farcs 2 \ (\sim 28 \ \mathrm{AU})$ scale at  
$0 \farcs 06 \ (\sim 8.4 \ \mathrm{AU})$ resolution.  
Interior to the sub-mm resolved gap, we find spiral structures, 
indicative of dynamical processes.
We use the spiral density wave theory to interpret the 
structure, and estimate disk's physical parameters.

 \section{Observations \& Data Reduction}

  \subsection {HiCIAO Observations}

  SAO 206462 was observed in the $H$-band ($1.6 \ \micron$) 
  using the high-contrast imaging instrument 
  HiCIAO \citep{tamu06,hodapp08,suzuki10} 
  on the  Subaru Telescope on 2011 May 20 UT as part of
  Strategic Explorations of Exoplanets and Disks with Subaru
  \citep[SEEDS,][]{tamu09}. 
  The adaptive optics system \citep[AO188;][]{haya04,minowa10}  
  provided a stable stellar point spread function 
  (PSF, $ \mathrm{FWHM} = 0 \farcs 06$). 
  We used a combined angular differential imaging (ADI) 
  and polarization differential imaging (PDI) mode with 
  a field of view of $10^{''} \times 20^{''}$ and a pixel scale of 
  $9.5 \ \mathrm{mas} \ \mathrm{pixel}^{-1}$.  
  A $0 \farcs 3$-diameter circular occulting mask was used 
  to suppress the bright stellar halo.  
  The half-wave plates were placed to four angular positions 
  from 0$^{\circ}$, 45$^{\circ}$, 22.5$^{\circ}$, and 67.5$^{\circ}$ in
  sequence with one $30 \ \mathrm{sec}$ exposure per wave plate
  position.  The total integration time of the
  $PI$ image was $780 \ \mathrm{sec}$ 
  after removing low quality images 
  with large FWHMs by careful inspections of the stellar PSF.

\subsection {PDI Data Reduction} 
\label{sec:PDIred}

  The raw images were corrected using IRAF\footnotemark[1]
  \footnotetext[1]{IRAF is distributed
  by the National Optical Astronomy Observatory, which is operated 
  by the Association of Universities for Research in Astronomy,
  Inc., under cooperative agreement with the National Science
  Foundation.}
  for dark current and flat-field following the standard reduction
  scheme.  We applied a distortion correction using globular cluster M5
  data taken within a few days, using IRAF packages GEOMAP and GEOTRAN.  
  Stokes $(Q,U)$ parameters and the $PI$ image were obtained in
  the standard approach 
  \citep[e.g.,][]{hink09} as follows.  
  By subtracting two images of extraordinary- and ordinary-rays 
  at each wave plate position,
  we obtained $+Q$, $-Q$, $+U$, and $-U$ images, 
  from which $2Q$ and $2U$ images were made 
  by another subtraction to eliminate remaining aberration.  
  $PI$ was then given by $PI = \sqrt{Q^{2} + U^{2}}$. 
   Instrumental polarization of HiCIAO at the Nasmyth instrument was
  corrected by following \citet{Joos08}. 
  From frame-by-frame deviations, the typical error of surface
  brightness (SB) was estimated to be 
  $\sim 5\%$ at $r \sim 0 \farcs 5$
  when averaged over $5 \times 5$ pixels ($\sim$ PSF scale).  
  Comparing different data reduction methods (frame selections and
  instrumental polarization estimates), we expect that the
  systematic uncertainty of the SB of $PI$ to be $\sim 10 \%$.

  \subsection {Contemporaneous Photometry}
  \label{contphot}

  Since SAO 206462 shows variability in NIR wavelengths \citep{sit11},
  it is important to take photometry simultaneously with disk
  observations. 
  $H$-band photometry was obtained just before and after
  the disk imaging without the coronagraphic spot with the adaptive
  optics, by sixteen $1.5 \ \mathrm{sec}$ exposures at four spatially
  dithered positions.  An ND10 filter ($9.8 \pm 0.1\%$ transmission) was
  used to avoid saturation.
  Using the MKO filter set, 
  the $H$-band  
  \citep[$\lambda_{\rm eff} = 1.615 \ \micron$, 
  $\mathrm{FWHM} = 0.29 \ \micron$,][]{T02} 
  magnitude was $6.96\pm0.07 \ \mathrm{mag}$. 

\begin{figure*}
 \plotone{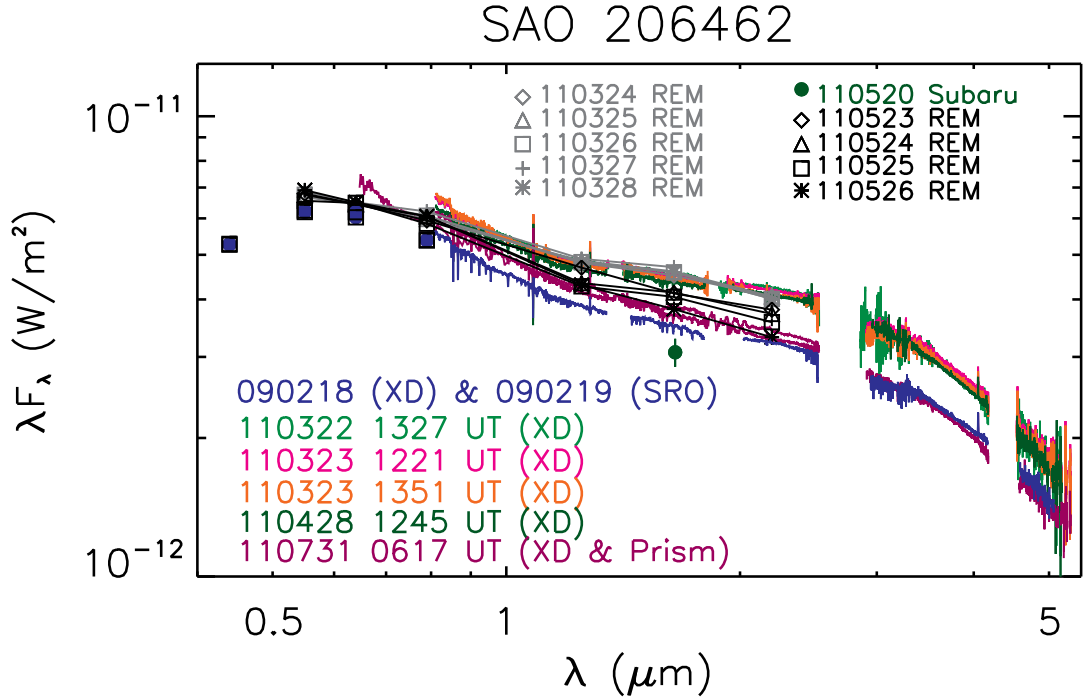}
 \caption{SED for SAO 206462 obtained by REM, adapted, in part, from
 \citet{sit11}.  The REM observations consist of \textit{VRIJHK} 
 photometry.
 Also shown are spectra obtained with the SpeX spectrograph.  
 See Section \ref{contphot} for the data reduction techniques.}
 \label{fig:REMSED}
\end{figure*}

  Broad-band \textit{VRIJHK} photometry was obtained on 2011 May 23-26, 
  starting within 48 hours of the HiCIAO observation,  
  using the Rapid Eye Mount (REM) Telescope at La Silla, Chile
  \citep{covino04}.  
  The REM $H$-band filter has 
  $\lambda_{\rm eff}=1.65 \ \micron$, $\mathrm{FWHM}=0.35 \ \micron$: 
  broader and displaced to longer wavelengths than the MKO filter.
  The observed data were reduced differentially using SAO 206463 (A0V). 
  The IR excess due to the inner disk 
  (Figure \ref{fig:REMSED}) was average for the range 
  observed in 2009-2011 \citep{sit11}.  
  No significant variation was observed 
  during the 2011 May observations, except for the small 
  long-term fading trend 
  ($\delta \mathrm{m} = 0.08 \pm 0.02 \ \mathrm{mag}$)  
  over the observation period.

  Figure \ref{fig:REMSED} also displays spectra obtained with the
  SpeX spectrograph \citep{rayner09} on NASA's Infrared Telescope
  Facility (IRTF).  
  The observations were obtained in the cross-dispersed (XD) echelle mode
  between $0.8$ and $5.1 \ \micron$ using a $0 \farcs 8$ 
  slit ($R \sim 900$) and calibrated using HD 129685 (A0V) with
   SpeXtool \citep{vacca03,cushing04}.
  The absolute flux calibration, to correct for
  light loss at the spectrograph slits, was accomplished in two
  ways: using photometry and wide-slit spectroscopy
  \citep[see][]{sit11}. 
  The March SpeX data were normalized 
  using the REM photometry, obtained in the days immediately after the
  SpeX observations, and when the star was photometrically stable. 
  In July, SAO 206462 and the calibration star were observed with the
  SpeX prism using a $3 \farcs 0$ slit, which, under good seeing and 
  transparency conditions, produces absolute flux to $\sim 5\%$ accuracy.  
  The REM photometry at $H$-band in May is $\sim 0.25 \ \mathrm{mag}$
  brighter than the Subaru data.   
   The Subaru photometry shows a low value even when considering
  the filter difference,
  suggesting that the outer disk is illuminated efficiently.

 \section{Results} 

  \subsection{Spiral Structure} 
  
  The SAO 206462 disk can be traced in $PI$ from
  $0 \farcs 2$ to $ \sim 1 \farcs 0$ 
  ($28-140 \ \mathrm{AU}$), similar to the range imaged  
  with HST/NICMOS \citep{gra09}, but with 
  a factor of $\sim 4$ greater angular resolution. 
  The total $PI$ is $9.87 \ \mathrm{mJy} \pm 0.06 \% $ 
  at $0 \farcs 2 < r < 1 \farcs 0$, 
  which is $0.6 \%$ of the stellar intensity.  
  The total $PI$ at $0 \farcs 42 < r < 1 \farcs 0$
  is $3.94 \ \mathrm{mJy} \pm 0.1 \% $ 
  while the total intensity by HST/NICMOS F110W is 
  $9.7 \ \mathrm{mJy}$ \citep{gra09}. 
  The average SB of $PI$ at $r=0 \farcs 46$ 
  is $\sim 6 \ \mathrm{mJy/asec}^2$, 
  whereas the total intensity by HST/NICMOS F160W is 
  $30 \ \mathrm{mJy/asec}^2$ \citep{gra09}. 
  Given the NICMOS data uncertainties, the polarization fraction is 
  $\sim 20-40\%$, assuming no PSF halo in the
  HiCIAO data and no variable self-shadowing/illumination in the disk.    
  Our measured polarization fraction is similar to that of 
  HD 100546 \citep[$14\%^{+19\%}_{-8\%}$,][]{qua11}
  and AB Aur \citep[$\sim 25 \% - 45 \%$,][]{P09}.  

\begin{figure*}
 \plotone{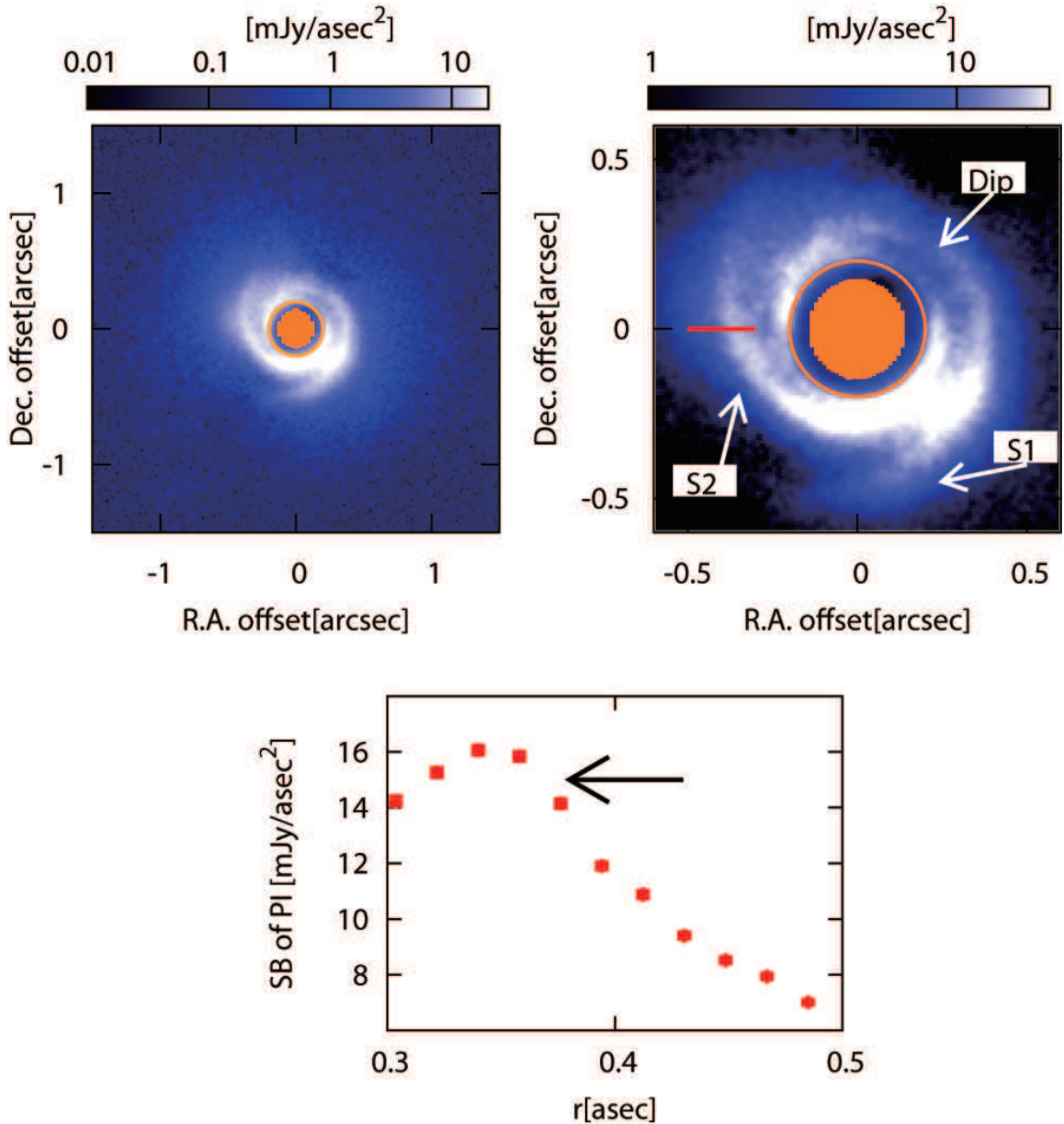}
 \caption{Top: $PI$ image of SAO 206462 in the 
 north-up configuration with log-stretch color scales. 
 The filled orange circles at the center indicate the mask size 
 ($r = 0 \farcs 15$).  The circles around have $r=0 \farcs 2$, exterior
 to which the features are considered to be real.  
 The right panel is central region's closeup.  Different color
 scales are used to enhance the spirals labeled as ``S1'' and ``S2''.  
 The ``Dip'' may be due to depolarization.    
 Bottom: $PI$ profile along the red line in the top right panel. 
 The arrow indicates the location of S2.   
 The position errors are not shown for visibility.}
 \label{fig:image}
\end{figure*}

   Figure \ref{fig:image} shows the $PI$ image.  
  The region interior to $0 \farcs 4$ 
  is not a void and  
  we do not see clear structural evidence of the cavity wall
  in \citet{and11} model 
  ($R_{\rm cav} = 46 \ \mathrm{AU} \sim 0 \farcs 33$).  
  We see spiral arcs S1 (east) and S2 (south-west).
  The $PI$ at the location of the spirals is $\sim 30 \%$ larger than 
  that extrapolated from the smooth outer profile 
  (bottom of Figure \ref{fig:image}).  
  The brightest portions of the spirals roughly coincide with 
  the bright thermal emission peaks at $12 \ \micron$ \citep{mar11} 
  and lie inside the ring noted by \citet{dou06}.
  It is also noted that we see a dip in $PI$ in the north-west, 
  probably due to the depolarization in the minor axis 
  direction (see below), and that we do not see large-scale, localized
  shadow that might be cast by the inner dust belt if highly inclined
  relative to the outer disk. 

  \subsection{Azimuthal and Radial Profiles}

\begin{figure*}
 \plotone{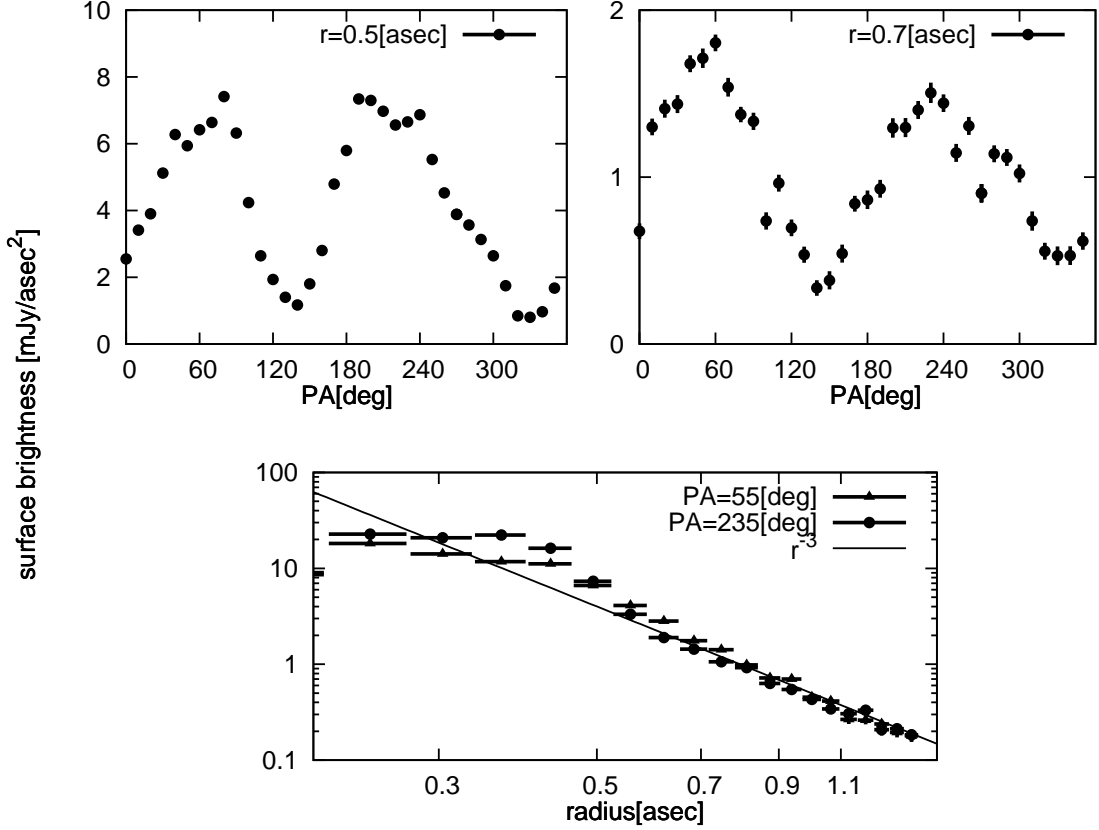}
 \caption{
 Top: Azimuthal SB profile at $r= 0 \farcs 5$ (left) 
 and $r= 0 \farcs 7$ (right) with PA measured from north to east.  
 Bottom: Radial profile along the major axis. 
 Position errors indicate the FWHM of the PSF.  SB errors are estimated
 using frame-by-frame deviations. 
 }
 \label{fig:profiles}
\end{figure*}

  Here, we summarize the overall disk structure exterior to the spirals.  
  Figure \ref{fig:profiles} (top panels) shows the azimuthal SB profiles
  at $r=0 \farcs 5$ and $r=0 \farcs 7$.  
  At $r \gtrsim 0 \farcs 5$, SB has 
  maxima around 
  $\mathrm{position\ angle\ (PA)} \sim 50^{\circ}-60^{\circ}$ and 
  $230^{\circ}-240^{\circ}$.  
  Since the polarization is maximized at 
  $\sim 90^{\circ}$ scattering \citep[e.g.,][]{GKM07}, 
  it is implied that the disk major axis is
  at $ \mathrm{PA} \sim 50^{\circ}-60^{\circ}$, 
  comparable to estimates by CO observations: 
  $\mathrm{PA}=56^{\circ} \pm 2^{\circ}$ by \citet{pon08} and 
   $64^{\circ} \pm 2^{\circ}$ by \citet{lyo11}.
  We adopt $\mathrm{PA}=55^{\circ}$ for the major axis and
  $i=11^{\circ}$ for inclination (see Section \ref{sec:intro}).  
  Our spiral model fitting results (the next section) 
  are little affected even if we assume a face-on geometry. 

  From CO observations, it is known that the south-west side is
  receding \citep[e.g.,][]{lyo11}.  
  Therefore, either the north-west or south-east side is
  the near side.  We do not see an obvious forward scattering
  excess in the NIR image. 
  However, since the spirals are typically trailing, 
  it is inferred that the south-east is the near side.

  Figure \ref{fig:profiles} also shows the radial $PI$ profiles
  along the major axis, which is roughly consistent
  with $r^{-3}$ in the outer part, 
  indicating a flat (not highly flared) disk \citep{WH92}.  
  The radial slopes vary as PA from $\sim -2$ to $\sim -4.5$ 
  (fitting at $0 \farcs 6 < r < 1 \farcs 0$) or  
  from $\sim -2.5$ to $\sim -5$ 
  (fitting at $0 \farcs 3 < r < 0 \farcs 9$), 
  with shallower slopes typically appearing in the minor axis
  directions.   
  However, $r^{-3}$ is representative on average.  
  This slope is observed in several other HAeBe disks
  \citep[e.g.,][for total intensity data]{F10}, 
  although HD 97048 disk exhibits a shallower slope \citep{Q11}.

  \section {Spiral Structure Modeling}
  \label{sec:spiral}

  Among several features in the image, the most interesting
  one is the non-axisymmetric spirals. 
  In order to understand them, 
  we propose a model based on the spiral density wave theory 
  \citep[e.g.,][]{LS64,GT78,GT79,OL02}, assuming that NIR emission traces
  the disk surface density structure. 
  With such a model, the spiral structures can be used to infer the disk
  temperature, independently of, for example, CO line observations.

  The shape of the spiral density wave is determined by the location of 
  the launching point (corotation radius $r_{\rm c}$) and disk's
  thermal and rotation profiles.  When the disk rotation angular
  frequency is $\Omega(r) \propto r^{-\alpha}$ and the sound speed is 
  $c(r) \propto r^{-\beta}$, 
  the shape of the wave far from $r_{\rm c}$ 
  is given by 
  \begin{eqnarray}
   \theta(r) &=& \theta_0 - \dfrac{\mathrm{sgn}(r-r_{\rm c})}{h_{\rm c}}
    \times
    \nonumber \\
   &&  \left[
	\left(\dfrac{r}{r_{\rm c}}\right)^{1+\beta} 
	\left\{ \dfrac{1}{1+\beta} - \dfrac{1}{1-\alpha+\beta}
	 \left( \dfrac{r}{r_{\rm c}} \right)^{-\alpha} \right\}
	- \left(\dfrac{1}{1+\beta} - \dfrac{1}{1-\alpha+\beta}\right)
	 \right]
   \label{spiralform}
  \end{eqnarray}
  in the polar coordinate $(r,\theta)$, 
  where $h_{\rm c} = c(r_{\rm c})/r_{\rm c}\Omega(r_{\rm c})$ 
  denotes the disk aspect ratio at $r=r_{\rm c}$ and $\theta_0$ gives the
  phase.  
  Equation \eqref{spiralform} well approximates 
  the shape of the density wave 
  given by the WKB theory \citep{R02,MSI10}.   
  When the spiral is excited by a planet 
  in a circular orbit, 
  its location is $\sim (r_{\rm c},\theta_0)$.  
  Equation \eqref{spiralform} has five parameters, 
  $(r_{\rm c}, \theta_0, h_{\rm c}, \alpha, \beta)$.

\begin{figure*}
 \plotone{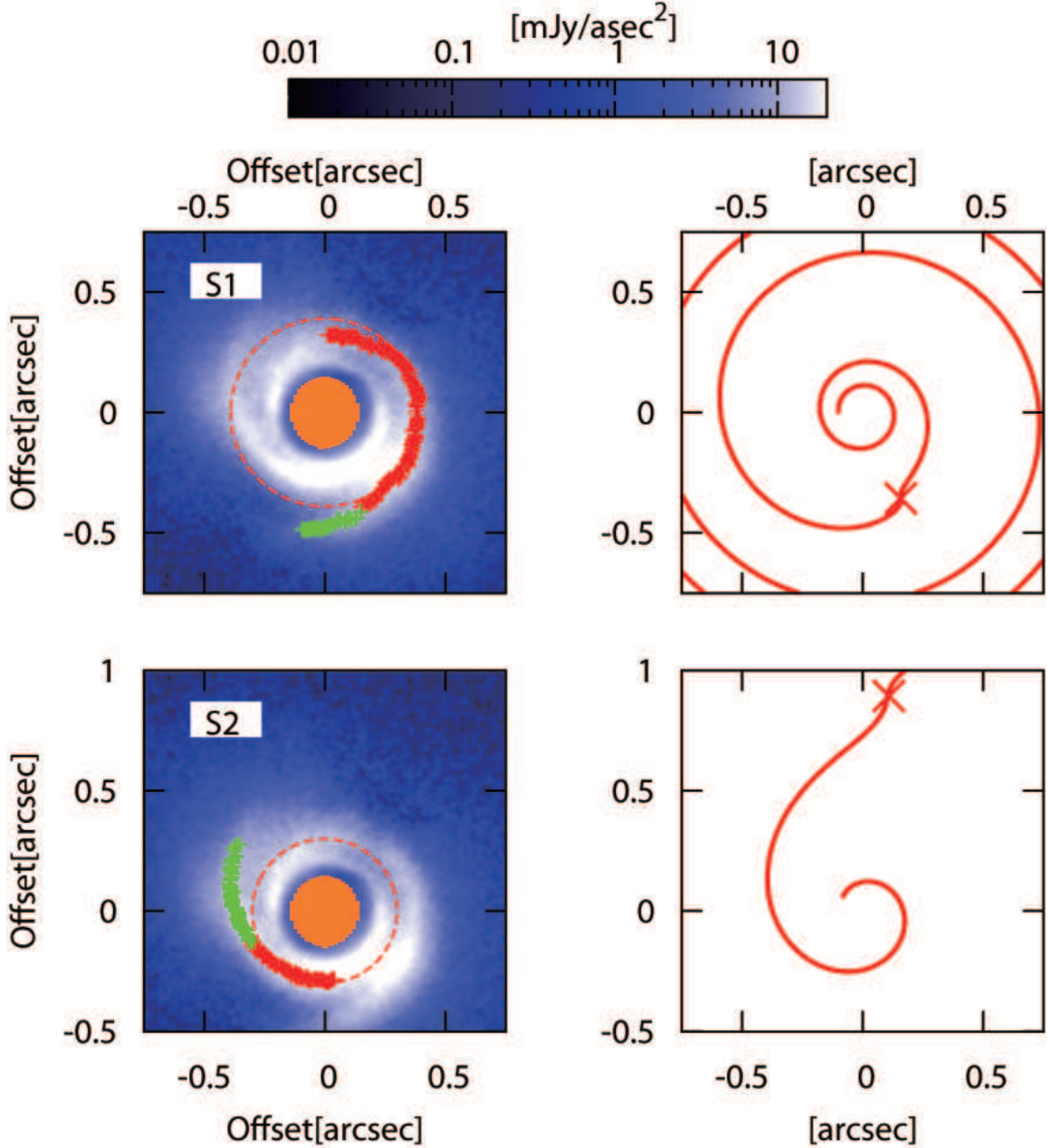}
 \caption{Left: Red and green points indicate the locations of the
 maxima of radial profiles for S1 (top) and S2 (bottom). 
 The green ones are used for the fitting by 
 Equation \eqref{spiralform} (Section \ref{sec:spiral}).  
 Dashed lines indicate circles with  
 $r=0 \farcs 39$ (top) and $r=0 \farcs 3$ (bottom).  
 Right: The spiral shape given by Equation \eqref{spiralform} with the
 best-fit parameters of S1 (top) and S2 (bottom).  Crosses show the
 locations of $(r_{\rm c},\theta_0)$.
 }
 \label{fig:maxima}
\end{figure*}

  Two non-axisymmetric features, S1 and S2, (Figure
  \ref{fig:maxima}) are identified as follows. First, 
  local maxima in the radial SB profiles normalized by $r^2$ 
  (to take into account the dilution of the stellar flux) 
  are traced at every $1^{\circ}$ step with data 
  at $170^{\circ}<\mathrm{PA}<360^{\circ}$ (S1) and at 
  $50^{\circ} < \mathrm{PA} < 190^{\circ}$ (S2). 
  The points near the minor axis 
  ($\mathrm{PA}\sim 325^{\circ}$) are excluded because 
  the structure there may be affected by depolarization.
  The points at $\mathrm{PA}>200^{\circ}$ (S1) 
  and $\mathrm{PA}>114^{\circ}$ (S2) 
  may be a part of axisymmetric rings since they
  are found to have constant radii.   
  After removing these points,   
  we have 27 (S1) and 56 (S2) points as representing samples of 
  non-axisymmetric spirals, with the opening angle of 
  $\sim 15^{\circ}$ for both S1 and S2. 
  We estimate that the uncertainty of the location of 
  the maxima is given by the FWHM of the PSF.  

  In order to fit the non-axisymmetric structures by Equation
  \eqref{spiralform},  we fix $\alpha$ and $\beta$ at $1.5$
  (Kepler rotation) and $0.4$, respectively, 
  as in \citet[][]{lyo11},
  while other parameters are varied as  
  $(0 \farcs 1<r_{\rm c}<0 \farcs 9, 
  0<\theta_0<2\pi,0.05<h_{\rm c}<0.25)$.   
  Note that different values of $\beta$ yield similar results.
  Since it is difficult to fit S1 and S2 simultaneously, they are
  fitted independently.   

\begin{figure*}
 \plotone{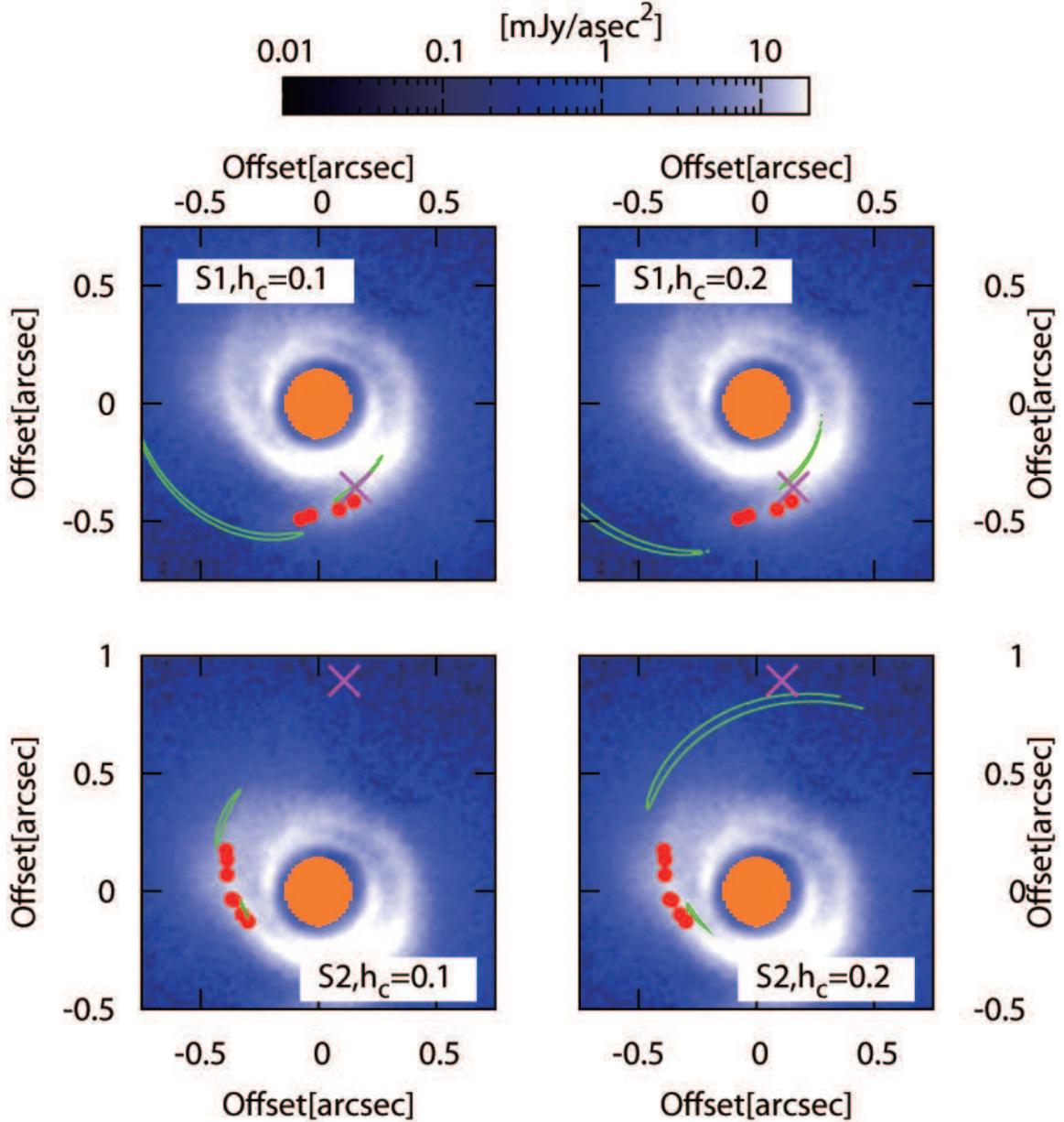}
 \caption{Parameter degeneracy of the fitting for S1 (top) 
 and S2 (bottom) with $h_{\rm c}=0.1$ (left) and 
 $h_{\rm c}=0.2$ (right).  The red points
 show a part of the data fitted.  The
 green curves indicate the $63.8\%$ confidence level for the locations
 of $(r_{\rm c},\theta_0)$.  The magenta crosses are 
 $(r_{\rm c},\theta_0)$ for the best-fit parameters.  Note
 that the best-fit values of $h_{\rm c}$ are different from those 
 shown in the figure.}
 \label{fig:parameterdegen}
\end{figure*}

  The ``best-fit'' parameters are 
  $(r_{\rm c},\theta_0,h_{\rm c})=(0 \farcs 39, 204^{\circ}, 0.08)$ 
  for S1 (reduced $\chi^2 = 0.52$) and 
  $(r_{\rm c},\theta_0,h_{\rm c})=(0 \farcs 9, 353^{\circ}, 0.24)$ 
  for S2 (reduced $\chi^2 = 0.31$). 
  The spiral shapes with these parameters are shown in
  Figure \ref{fig:maxima}.  
  However, the parameter degeneracy is significant. 
  Figure \ref{fig:parameterdegen} shows the parameter space
  of $(r_{\rm c},\theta_0)$ with $63.8\%$ confidence level 
  for $h_{\rm c}=0.1$ and $h_{\rm c}=0.2$.  
  Note that in Figure \ref{fig:parameterdegen}, 
  the ``best-fit'' of $(r_{\rm c},\theta_0)$ 
  is outside the domain of confidence in some cases 
  because $h_{\rm c}$ is not the same as the best-fit.  
  Despite the parameter degeneracy, the values of the aspect ratio 
  which fit the shape of the spiral ($h_{\rm c}\sim 0.1$) 
  are consistent with those obtained from the sub-mm map of the disk 
  \citep[e.g., 
  $h=0.096 (r/100 \mathrm{AU})^{0.15}$;][]{and11}.

  The spiral density wave theory predicts that  
  the pattern speed deviates from the local Kepler speed;
  \begin{equation}
   \Omega_{\rm pattern} = 0.8
    \left( \dfrac{r_{\rm c}}{70\mathrm{AU}} \right)^{-3/2}
    \left( \dfrac{M_{\ast}}{1.7M_{\odot}} \right)^{1/2}
    [\mathrm{deg/yr}]
  \end{equation}
  is not necessarily equal to $\Omega(r)$.  
  When $r_{\rm c}=0 \farcs 5 (\sim 70\mathrm{AU})$, the spiral will move
  $\sim 10^{\circ}$ in a decade, 
  corresponding to movement of $0 \farcs 1$.  
  Considering the PSF scale of our observations and the locations of the
  spirals, such deviations can be detectable over a couple of decades.    
  Moreover, if the two spirals have distinct corotation radii, 
  their relative locations change in time due to the pattern speed 
  difference.  
  Such measurements will confirm that the observed feature is really the
  density wave, providing indisputable evidence of dynamical activity.  

  Note that it would be difficult to detect spirals in colder
  disks (smaller $h_{\rm c}$), where spirals are more tightly wound, 
  due to the blurring by the PSF.
  The lower detectable limit of $h_{\rm c}$ is typically
  $h_{\rm c} \sim 0.01 - 0.03$ for our set of parameters. 
  The combination of high angular resolution and warm temperatures allows 
  the spiral structure in the SAO 206462 disk to be resolved. 
  Further spirals might be detectable in similarly warm disks.

 \section{Summary and Discussion}

 In this Letter, we present a high-resolution image of the
 SAO 206462 transitional disk using Subaru/HiCIAO, with an inner working
 angle of $0 \farcs 2$. 
 We discover non-axisymmetric spiral features, which can be explained by
 the spiral density wave theory with a reasonable value of the disk
 aspect ratio ($h_{\rm c} \sim 0.1$).
 The model is robust in a sense that it does not assume the 
 origin of such structure explicitly.  

 The detection of scattered light within the sub-mm cavity itself is
 interesting, since \citet{and11} predicts that the sub-mm cavity is
 heavily depleted.   
 Our data in tandem with the millimeter data may suggest that the
 depletion of grains at different sizes is not uniform.
\citet{D12} discuss such discrepancies between sub-mm dust continuum
images and NIR scattered light images 
in terms of differing spatial distributions as a 
function of grain size from a general theoretical perspective.

 Our major assumption is that $PI$ at $H$-band, tracing the scattered
 light at the disk surface,  
 actually traces the surface density variations.  
 This assumption is valid when the disk is in vertical, isothermal
 hydrostatic equilibrium without rapid radial surface density
 variations \citep[e.g.,][]{Muto11}. 
 Structures near the midplane are, however, preferentially observed 
 at longer wavelengths with high spatial resolution: 
 Atacama Large Millimeter/Submillimeter Array (ALMA) can be an ideal
 instrument.

 Among several possible causes for the spiral structures 
 \citep[see also][]{hash11},  
 one interesting idea is that planets excite them. 
 In this case, the domain of possible locations of the planets is given
 by the green curves in Figure \ref{fig:parameterdegen}.
 If the two spirals have distinct corotation radii, there may be two
 (unseen) planets embedded in the disk. 
 The amplitude of the surface density perturbation scales with the planet 
 mass as $\delta \Sigma / \Sigma \sim GM_{\rm p}\Omega/c^3$
 for non-gap-opening low-mass planets   
 \citep[$GM_{\rm p}\Omega/c^3 \lesssim 1$, e.g., ][]{TTW02}. 
 In our data, the amplitude of the spiral wave is 
 typically $\sim 30\%$ (Figure \ref{fig:image}), 
 implying $M_{\rm p} \sim 0.5 M_{\rm J}$.  

 The typical error of $PI$ in our image is $\sim 5 \%$.  
 If this is typical of HiCIAO, it is capable of detecting the indirect
 signatures of planets down to $M_{\rm p} \sim 0.05M_{\rm J}$.  
 ADI is promising in finding a point source; 
 however, small field rotation ($\sim 13^{\circ}$) due to the
 southerly declination ($-39^{\circ}$) of SAO 206462 
 makes obtaining the total intensity difficult in our data. 
 Future $L$-band observations may reveal thermal emission from a
 planet, if it exists, or its surrounding (accreting) gas.   

 \acknowledgments 
 The authors thank Roman Rafikov for comments, the support
 staff members of the IRTF and REM telescopes for assistance in
 obtaining the SED data, and the IR\&D program at The Aerospace
 Corporation.  
 REM data in this study were obtained under Chilean National TAC
 programs CN2011A-050 and CN2011B-31.
 This work is partially supported by KAKENHI 22000005 (MT), 
 23103002 (MH and MH), 23103004 (MM and MF), 23103005, 
 23244027, 18540238 (SI), and 22$\cdot$2942 (TM), 
 WPI Initiative, MEXT, Japan (ELT), 
 NSF AST 1008440 (CAG), 1009203 (JC), and 1009314 (JPW), 
 and NASA NNH06CC28C (MLS) and NNX09AC73G (CAG and MLS).
 Part of this research was carried out at JPL.

\end{document}